\documentclass[aps,prl,showpacs,twocolumn]{revtex4}

\usepackage{amssymb}
\usepackage{epsfig}
\usepackage{graphicx}
\usepackage{amsmath}
\usepackage{array,color}
\usepackage{dcolumn}
\usepackage{bm}

\begin{document}

\title{Kondo Metal and Ferrimagnetic Insulator on the Triangular Kagom\'e Lattice}
\author{Yao-Hua Chen$^{1}$}
\author{Hong-Shuai Tao$^{1}$}
\author{Dao-Xin Yao$^{2}$}\email{yaodaox@mail.sysu.edu.cn}
\author{Wu-Ming Liu$^{1}$}\email{wmliu@iphy.ac.cn}
\affiliation{
$^1$Beijing National Laboratory for Condensed Matter Physics, Institute of Physics, Chinese Academy of Sciences, Beijing 100190, China \\
$^2$State Key Laboratory of Optoelectronic Materials and
Technologies,Sun Yat-sen University, Guangzhou 510275, China}


\date{\today}

\begin{abstract}

We obtain the rich phase diagrams in the Hubbard model on the
triangular Kagom\'e lattice as a function of interaction,
temperature and asymmetry, by combining the cellular dynamical
mean-field theory with the continuous time quantum Monte Carlo
method. The phase diagrams show the asymmetry separates the
critical points in Mott transition of two sublattices on the
triangular Kagom\'e lattice and produces two novel phases called
plaquette insulator with an obvious gap and a gapless Kondo metal.
When the Coulomb interaction is stronger than the critical value
$U_c$, a short range paramagnetic insulating phase, which is a
candidate for the short rang resonating valence-bond spin liquid,
emerges before the ferrimagnetic order is formed independent of
asymmetry. Furthermore, we discuss how to measure these phases in
future experiments.

\end{abstract}


\maketitle


Geometrically frustrated systems have shown a lot of interesting
phenomena, such as the spin liquid and spin ice \cite{Meng,
Bramwell, Mengotti, LEESH, Sondhi, Wang}. The charge and magnetic
order driven by interaction is still one of the central issues in
the field of strongly correlation systems \cite{Loh, Takuya,
Yasuyuki, Takuma, Jarrell63, wu}. Recently, a new class of
two-dimensional materials $Cu_9X_2(cpa)_6\cdot xH_2O$
(cpa=2-carboxypentonic acid, a derivative of ascorbic acid; X=F,
Cl, Br) has been found \cite{Gonzalez,Maruti,Mekata}, which is
formed by an extra set of triangles (B-sites in Fig. 1) insides of
the Kagom\'e triangles (A-sites in Fig. 1). The Cu spins form a
new type of geometrically frustrated lattice called triangular
Kagom\'e lattice (TKL). This lattice can also be realized by cold
atoms in the optical lattices, in which the interaction between
trapped atoms can be tuned by Feshbach resonance and the kinetic
energy can be adjusted by the lattice depth \cite{Jaksch,
Immanuel, Hofstetter, Chen1, Duan, Panahi, Gemelke}. Although the
effective spin models on this system have been investigated
\cite{Loh, Yamada}, the real charge dynamics with spins and the
phase diagram on this lattice have not been studied. Different
with other frustrated system, such as the triangular lattice and
Kagom\'e lattice, the ``triangles-in-triangles'' structure on the
TKL induces two different sublattice. Therefore, it is desirable
to investigate the Mott and magnetic transition on the TKL under
the influence of asymmetry which is induced by different hoppings
between two sublattices.

\begin{figure}[t]
 \centering
\includegraphics[width=12.0cm]{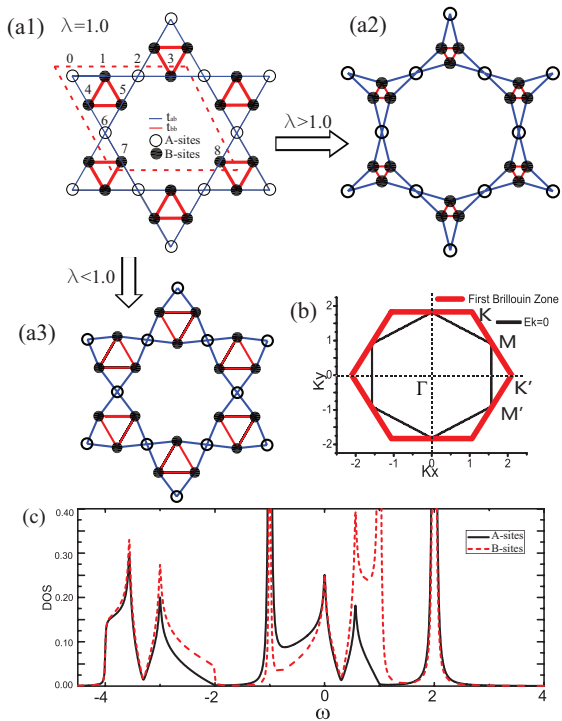}\hspace{0.5cm}
\caption{\label{fig:epsart}(Color online) (a1) The unit cell of
the triangular Kagom\'e lattice (TKL) without asymmetry
($\lambda=1.0$). The open circles denote A-sites and the solid
circles denote B-sites. The blue lines represent the hopping
between A-sites and B-sites. And the red lines denote the hopping
between B-sites. (a2) When $\lambda>1.0$, the TKL is similar with
the Kagom\'e lattice. (a3) When $\lambda<1.0$, the TKL is
transformed into a system composed of many triangular plaquettes.
(b) The thick red lines show the first Brillouin zone of the
triangular Kagom\'e lattice. The thin black lines correspond to
the Fermi surface for the non-interacting case. The $\Gamma, K, M,
K', M'$ points denote the points in first Brillouin zone with
different symmetry. (c) The density of states of the triangular
Kagom\'e lattice for the A- and B-sites in the case of $U=0$ and
$\lambda=t_{ab}/t_{bb}=1.0$.}
\end{figure}

Recently, many analytical and numerical methods have been
developed to investigate the strongly correlated system
\cite{Dimitrios,  Aryanpour, niu, CCC, xiaod, Hlee, Terletska,
Fzhou}, such as the dynamical mean-field theory (DMFT)
\cite{Antoine}. However, the DMFT works ineffectively in the
frustrated systems because the nonlocal correlation and spatial
fluctuations are ignored. Therefore, a new method called cellular
dynamical mean-field theory (CDMFT) has been developed to
incorporate spatially extended correlations and geometrical
frustration in the framework of DMFT \cite{Thomas, Parcollet,
Lorenzo, Park} by mapping the original lattice problem onto an
effective cluster model coupled to an effective medium. The
continuous time quantum Monte Carlo method (CTQMC) \cite{Rubtsov}
is employed as an impurity solver in the CDMFT loop, which is more
accurate than the translational QMC method due to the absence of
Trotter decomposition. So, the CDMFT combining with the CTQMC can
provide useful numerous insights into the phase transitions in the
frustrated system \cite{Takuya, Park}.

\begin{figure}[t]
 \centering
\includegraphics[width=8.0cm]{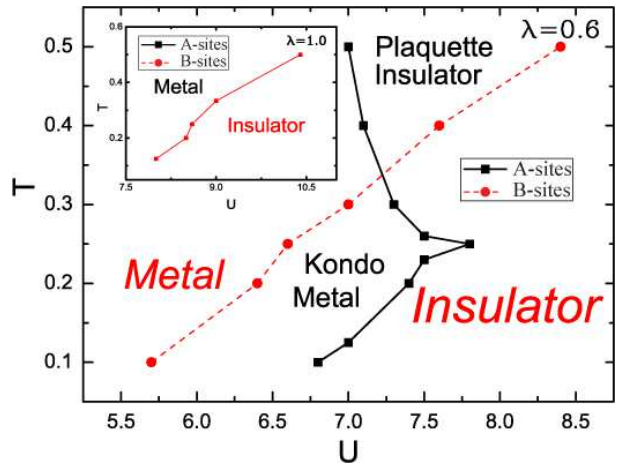}\hspace{0.5cm}
\caption{\label{fig:epsart}(Color online) Phase diagram of the
triangular Kagom\'e lattice at $\lambda=0.6$ with $t_{bb}=1.0$ as
the unit of energy. The black solid lines with square points show
the transition line of the A-sites, and the red dashed lines with
circle points show the transition line of the B-sites. Two kinds
of coexisting phases between red lines and black lines are the
plaquette insulating phase and Kondo metal phase. Inset: The phase
diagram of the symmetric TKL, i.e. $\lambda=1$, in which there are
no coexisting phases.}
\end{figure}

In the present Letter, we employ the CDMFT combining with the
CTQMC to investigate the influence of asymmetry on the Mott and
magnetic phase transition on the TKL based on Hubbard model. Two
novel phases called plaquette insulator and Kondo metal induced by
asymmetry is found in the phase diagram and their properties is
studied by determining the momentum resolved spectrums. By
defining a magnetic order parameter, we can characterize a
ferrimagnetic order for the large interaction on the TKL. The
phase diagram with the competition between interaction and
asymmetry is provided. In addition, the spectrum functions on the
Fermi surface for the different interaction and asymmetry are
presented. These interesting phases can be probed by the
angle-resolved photoemission spectroscopy (ARPES)
\cite{Damascelli}, neutron scattering, nuclear magnetic resonance
(NMR) and other experiments.

We consider the standard Hubbard model on the TKL,
\begin{equation}\label{eq:eps}
H=-\sum_{<ij>\sigma}t_{ij}c_{i\sigma}^{+}c_{j\sigma}+U\sum_{i}n_{i\uparrow}n_{i\downarrow}+\mu\sum_{i\sigma}n_{i\sigma},
\end{equation}
where $t_{ij}$ is the nearest-neighbor hopping energy, $U$ is the
Coulomb interaction, $\mu$ is the chemical potential,
$c_{i\sigma}^{+}$ and $c_{i\sigma}$ denote the creation and the
annihilation operators, and
$n_{i\sigma}=c_{i\sigma}^{+}c_{i\sigma}$ corresponds to the
density operator. The asymmetry factor is defined as
$\lambda=t_{ab}/t_{bb}$, which can be adjusted by suppressing the
samples in experiments. For the convenience, we use $t_{bb}=1.0$
as the energy unit. As shown in Fig. 1(a), there are two different
sublattices on the TKL called A-sites and B-sites. The space group
of the TKL is $p6m$ same as the honeycomb lattice in the case of
$\lambda=1$, as show in Fig. 1(a1). Fig. 1(a2) shows the TKL is
similar as the Kagom\'e lattice when $\lambda>1$. We find the TKL
can be transformed into a system composed of many triangular
plaquettes in the case of $\lambda<1$, as shown in Fig. 1(a3). The
first Brillouin zone and Fermi surface of the symmetric TKL
($\lambda=1$) with $U=0$ is shown in Fig. 1(b). The density of
states (DOS) for the different sublattices are shown in Fig. 1(c)
in the case of $U=0$ and $\lambda=1$. They have the similar DOS
around the Fermi surface.

We improve the CDMFT by combining it with the CTQMC to investigate
the correlation effects on the TKL. In the CDMFT, the original
lattice is mapped onto an effective cluster model via a standard
DMFT procedure. Thus, the effective medium $\hat{g}$ can be
obtained by a Dyson equation:
\begin{eqnarray}
\hat{g}^{-1}(i\omega)=(\sum_{\vec{k}}\frac{1}{i\omega+\mu-\hat{t}(\vec{k})-\hat{\Sigma}(i\omega)})^{-1}+\hat{\Sigma}(i\omega),
\end{eqnarray}
where $\hat{t}(\vec{k}$) is the hopping matrix element in the
cluster, $\vec{k}$ is the wave vector within the reduced
Brioullion zone based on clusters, $\hat{\Sigma}(i\omega)$ is the
self-energy matrix, and $\omega$ is the Matsubara frequency. Then,
we introduce an impurity solver, such as the CTQMC, to calculate
the cluster Green function $\hat{G}(i\omega)$. The new self-energy
$\hat{\Sigma}(i\omega)$ can be calculated by Dyson equation
$\hat{\Sigma}(i\omega)=\hat{g}^{-1}(i\omega)-\hat{G}^{-1}(i\omega)$
to close the self-consistent iterative loop. This CDMFT loop is
repeated until the $\hat{\Sigma}(i\omega)$ converges to a desired
accuracy. The CDMFT contains the important information of nonlocal
correlation and geometrical frustration, which is absent in the
general dynamical mean-field theory (DMFT). In addition, the
absence of Trotter decomposition in the CTQMC makes this method
more accurate than the translational auxiliary-field QMC.

 \begin{figure}[t]
 \centering
\includegraphics[width=8.0cm]{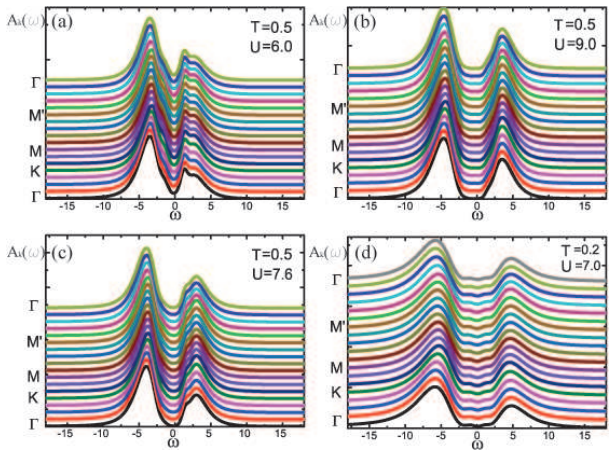}\hspace{0.5cm}
\caption{\label{fig:epsart}(Color online) The momentum resolved
one-particle spectrum $A_k(\omega)$ at $\lambda=0.6$ in the units
of $t_{bb}=1$. (a) The metallic phase in the weak interaction
case, such as $U=6.0$ and $T=0.5$. (b) The Mott insulating phase
in the strong interaction case, such as $U=9.0$ and $T=0.5$. An
obvious single particle gap shows up around the Fermi energy. (c)
The plaquatte insulating phase in the intermediate interaction
case, in which the A-sites are insulating but B-sites are
metallic, such as $U=7.6$ and $T=0.5$. A small gap shows up in
this phase. (d) The Kondo metallic phase at $U=7.0$ and $T=0.2$,
in which A-sites are metallic, B-sites are insulating. The single
particle gap vanishes in this phase.}
\end{figure}


The phase diagram obtained at $\lambda=0.6$ shows two coexisting
phases and a reentrant behavior of the Mott transition when
half-filling. The asymmetry can cause the separation of the phase
transition points of the A- and B-sites. As shown in Fig. 2, with
an increasing interaction at $T=0.5$, the A-sites are found to
translate from a metallic phase into an insulating phase and the
B-sites stay in the insulating phase. This phase is called the
plaquette insulating phase shown in Fig. 2. In this plaquette
insulating phase, electrons are localized on the A-sites and the
absence of next nearest neighbor hoppings causes the electrons can
only be itinerant within the B-sites. When $T<0.34$, the B-sites
translates into the insulating phase, however the A-sites remain
in the metallic phase, see Fig. 2. This coexisting phase
corresponds to a Kondo metal. The localized electrons on the
B-sites act as the magnetic impurity, and the electrons on the
A-sites are still highly itinerant. In this phase, the system
shows the strong Kondo effect because of the high density of
magnetic impurities. When the $U>U_c$, for example, $U_c=7.6$ at
$T=0.25$, both the A-sites and B-sites translate into insulators.
In Fig. 2, we find a reentrant behavior in the Mott transition of
the A-sites caused by the frustration and the asymmetry, which is
also found in the anisotropic triangular lattice \cite{Takuma}.
This reentrant behavior divides the coexisting phases into two
parts: the plaquette insulating phase and Kondo metallic phase.
The inset figure in Fig. 2 shows the phase diagram of the TKL at
$\lambda=1.0$. There are no coexisting phase and reentrant
behavior found on the TKL when the asymmetry is absent, i.e.
$\lambda=1$.

\begin{figure}[t]
 \centering
\includegraphics[width=7.0cm]{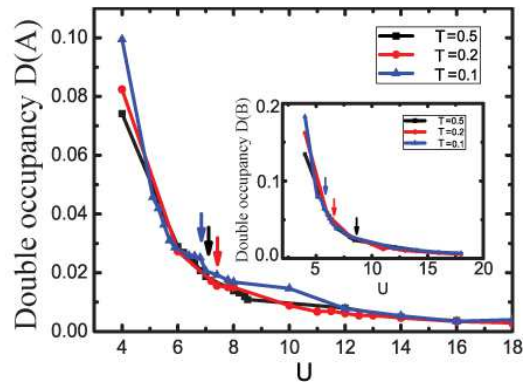}\hspace{0.5cm}
\caption{\label{fig:epsart}(Color online) The evolution of double
occupancy D on the A-sites as a function of interaction $U$ with
different temperatures at $\lambda=0.6$ in the units of
$t_{bb}=1$. The inset figure shows the evolution of double
occupancy on the B-sites. The arrows with different colors show
the phase transition points at different temperatures.}
\end{figure}

In order to find the properties of the different phases in Fig. 2,
we investigate the momentum resolved one-particle spectrum
$A_k(\omega)$ by using the Maximum Entropy Method (MEM)
\cite{Gubernatis}. The results are shown in Fig. 3. The $\Gamma,
K, M$ and $M'$ points have been defined in Fig. 1 (b). In the case
of $T=0.5$ and $U=6.0$, we find two obvious quasi-particles near
$\omega=4.0$ and $\omega=2.0$, see Fig. 3 (a). The gapless
behavior shows that the system stays in the metallic phase shown
in Fig. 2. When the interaction U increases,  we find an obvious
gap near the Fermi energy (see Fig. 3 (b)), which indicates the
system becomes an insulator denoted in Fig. 2. $A_k(\omega)$ of
the plaquette insulating phase is shown in Fig. 3 (c). In Fig. 3
(d), we find there is no obvious gap near the Fermi energy when
the system stays at the Kondo metal state in Fig. 2. The momentum
resolved one-particle spectrum function can be obtained by the
ARPES experiment, and the plaquette insulator and Kondo metal
phases can be found in certain samples. We define the double
occupancy $D$ on the TKL as
\begin{equation}\label{eq:eps}
D=\partial{F}/\partial{U}=\frac{1}{N}\sum_{i}\langle{n_{i\uparrow}n_{i\downarrow}}\rangle.
\end{equation}
The decreasing of $D$ on both sublattices shows that the
suppressing of the itinerancy of electrons, which is a
characteristic behavior of the Mott transition. When temperature
drops, D increases due to the enhancing of the spin fluctuation at
low temperature. The arrows indicate the phase transition point at
different temperatures. The smoothly decreasing $D$ characterizes
a second order Mott transition when the interaction $U$ increases.

\begin{figure}[t]
 \centering
\includegraphics[width=9.0cm]{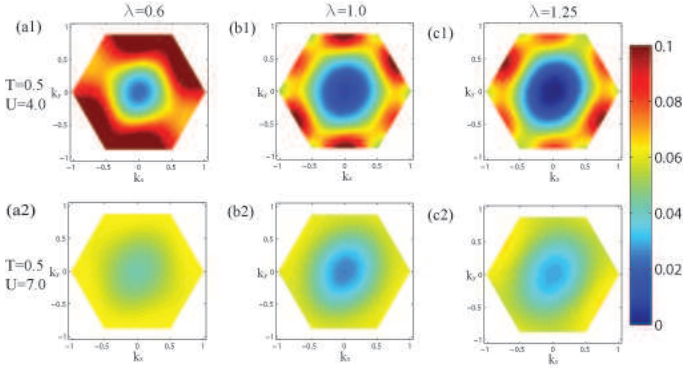}\hspace{0.5cm}
\caption{\label{fig:epsart}(Color online) The evolution of the
spectrum function on Fermi surface in the units of $t_{bb}=1$. (a)
$\lambda=0.6$, (b) $\lambda=1.0$, (c) $\lambda=1.25$.}
\end{figure}

In Fig. 5, we show the evolution of spectrum function on Fermi
surface as a function of $\vec{k}$ for different interactions $U$,
temperature $T$ and asymmetry $\lambda$, which is defined as
\begin{equation}\label{eq:eps}
A(\textbf{k};\omega=0)\approx-\lim_{\omega\rightarrow0}ImG_k(\omega+i0)/\pi,
\end{equation}
where
\begin{equation}\label{eq:eps}
G_k(\omega)=\frac{1}{N}\sum_{\gamma,\delta}e^{i\textbf{k}\cdot(\textbf{r}_{\gamma}-\textbf{r}_{\delta})}[\omega+\mu-\hat{t}(\textbf{k})-\hat{\Sigma}(\omega)]^{-1}_{\gamma\delta}.
\end{equation}
and $k$ is the wave vector in the original Brillouin zone. A
linear extrapolation of the first two Matsubara frequencies is
used to estimate the self-energy to zero frequency
\cite{Parcollet}. In the case of $\lambda=1.0$, the spectrum
function on Fermi surface shows six points near the M point in
Fig. 1 (b), similar as the non-interacting case. When interaction
$U$ increases, the Fermi surface shrinks because of the
localization of electrons. Fig. 5 (a1) shows the Fermi surface is
similar as the system composed of many triangular plaquettes at
$\lambda=0.6$, due to the suppression of the hopping between the
A-sites and the B-sites. The Fermi surfaces at $\lambda=1.0$ and
$\lambda=1.25$ are similar as a Kagom\'e lattice. When the
interaction $U$ keeps increasing, the Fermi surface is developing
into a flat plane showing the localization of electrons, see Fig.
5 (a2), (b2), and (c2). The spectrum function can be measured by
the ARPES experiment.

\begin{figure}[t]
 \centering
\includegraphics[width=8.0cm]{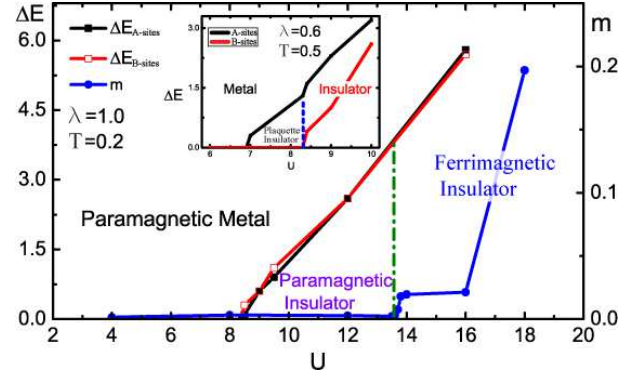}\hspace{0.5cm}
\caption{\label{fig:epsart}(Color online) The evolution of single
particle gap $\Delta{E}$ and ferrimagnetic order parameter $m$ at
$\lambda=1.0$ and $T=0.2$ in the units of $t_{bb}=1.0$. A
paramagnetic metallic phase is found when the interaction is weak
with $\Delta{E}=0$ and $m=0$. As the interaction $U$ increases, a
gap is opened and no magnetic order is formed with
$\Delta{E}\neq0$ and $m=0$. This paramagnetic insulating phase can
be a short range RVB spin liquid. An obvious magnetic order is
formed when the interaction is strong enough with $\Delta{E}\neq0$
and $m\neq0$. The insert picture shows the evolution of
$\Delta{E}$ at $\lambda=0.6$ and $T=0.5$. A plaquette insulator is
found when the A-sites are insulating and the B-sites are
metallic.}
\end{figure}

In order to investigate the evolution of magnetic order on the TKL,
we define a ferrimagnetic order parameter by:
\begin{equation}\label{eq:eps}
m=\frac{1}{N}\sum_{i}sign(i)(\langle{n_{i\uparrow}}\rangle-\langle{n_{i\downarrow}}\rangle),
\end{equation}
where $sign(i)=+1$, for i=0, 2, 6; and $sign(i)=-1$, for i=1, 3,
4, 5, 7, 8. Fig. 6 shows the evolution of the single particle gap
$\Delta{E}$ and m as a function of interaction $U$ in the case of
$\lambda=1.0$, $T=0.2$. The Mott transition points of the A-sites
and B-sites coexist when the asymmetry is absent. A gap opens in
the case of $U=8.5$. The ferrimagnetic order parameter $m=0$ at $U
< U_c=13.8$ indicates the system is in a paramagnetic insulating
phase. We argue that this paramagnetic insulating phase is a
candidate for the short range RVB spin liquid due to the absence
of long range correlations. When $U>U_c$, the finite $m$ means the
system enters a ferrimagnetic state. The inset picture of Fig. 6
shows the evolution of single particle gap $\Delta{E}$ at
$\lambda=0.6$ and $T=0.5$, in which a gap opens on the A-sites
first.

Finally, we provide a phase diagram about the asymmetry $\lambda$
and interaction $U$ at $T=0.2$, see Fig. 7. As the asymmetry
factor $\lambda$ increases, a coexisting zone containing the
plaquette insulator and the Kondo metal shows up. This zone is
suppressed from $\lambda=0.9$ to $\lambda=1.11$. When $U>U_c$,
such as $U_c=13.8$ in the case of $\lambda=1.0$, the system
becomes a ferrimagnetic insulator with $m\neq0$. Before entering
the ferrimagnetic phase, a paramagnetic insulator state is found.
We argue this paramagnetic insulator state is a candidate of a
short range RVB spin liquid due to the absence of any magnetic
order and long range correlations.

\begin{figure}[t]
 \centering
\includegraphics[width=8.0cm]{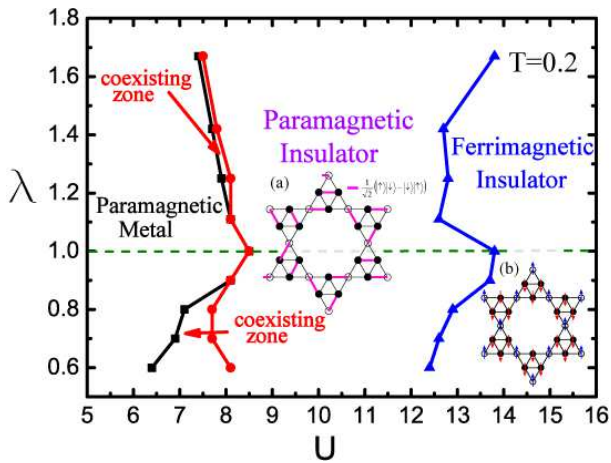}\hspace{0.5cm}
\caption{\label{fig:epsart}(Color online) The phase diagram of the
triangular Kagom\'e lattice represents the competition between
interaction and asymmetry for $T=0.2$ in the units of $t_{bb}=1$.
The region between the black lines with square points and the red
lines with circle points denotes the coexisting zone which
contains the plaquette insulator and the Kondo metal parts. A wide
paramagnetic insulating region is found with an intermediate $U$.
The blue lines with triangular points show the transition point to
the ferrimagnetic insulator with a clear magnetic order. Insert:
(a) Dimers formed in the paramagnetic insulator, which is a
candidate for the short range RVB spin liquid, (b) Spin
configuration of ferrimagnetic insulator.}
\end{figure}


In summary, we have obtained the rich phase diagrams as the
function of interaction U, temperature T, and asymmetry factor
$\lambda$. The asymmetry introduced in this strongly frustrated
system can change the shape of the metal-insulator translation
line and induces several interesting phases, such as plaquette
insulator and Kondo metal. In the plaquette insulator phase, an
obvious gap is found near the Fermi energy by investigating the
momentum resolved spectrum function. The Kondo metal phase is an
interesting phase where the partial sites work as the magnetic
impurities in the metallic environment. We also find an
interesting paramagnetic insulating phase which is related to the
short range RVB spin liquid state before the ferrimagnetic order
is formed. And the increasing of asymmetry does not suppress the
emerging of this ferrimagnetic order. In addition, we have
presented all kind of spectrum which can be used to detect these
phases in real materials, such as $Cu_9X_2(cpa)_6\cdot xH_2O$,
under applied pressure in experiments. Our studies provide a
helpful step for understanding the coexisting behavior in
metal-insulator transition, the formation of magnetic order and
the emerging of spin liquid in frustrated systems with asymmetry.

We would like to thank N. H. Tong and S. Q. Shen for valuable
discussions. This work was supported by the NKBRSFC under grants
Nos. 2011CB921502, 2012CB821305, 2009CB930701, 2010CB922904, and
NSFC under grants Nos. 10934010, 60978019, 11074310, and NSFC-RGC
under grants Nos. 11061160490 and 1386-N-HKU748/10, and RFDPHE of
China (20110171110026).

\end{document}